\documentclass{moriond}

\usepackage{amsmath}
\usepackage{amssymb}

\def\be{\begin{equation}}
\def\ee{\end{equation}}
\def\bea{\begin{eqnarray}}
\def\eea{\end{eqnarray}}

\newcommand{\Lya}{Ly-$\alpha$ }

\newcommand{\units}[1]{{\rm\ #1}}

\begin{document}
\vspace*{4cm}
\title{Mitigating Astrophysical Uncertainties in 21-cm Cosmology}

\author{ Omer Zvi Katz }

\address{School of Physics and Astronomy, Tel-Aviv University, Tel-Aviv 69978, Israel}

\maketitle


\abstracts{The light of the first astrophysical objects is expected to leave an imprint on the global 21-cm signal as it heats, excites, and ionizes neutral hydrogen. This dependence on early astrophysics introduces significant uncertainties in modeling the 21-cm signal during Cosmic Dawn (CD). Here we show that a combination of observables including high-redshift UV luminosity functions, the cosmic X-ray background, the optical depth to reionization, and hydrogen absorption lines in quasar spectra, can be used to mitigate the astrophysical uncertainties assuming minimal modeling. Beyond its implications to standard astrophysics, we demonstrate how applying this procedure can improve sensitivity to new physics signatures in the global 21-cm signal. Taking the scenario of fractional millicharged dark matter (DM) as an example, we address astrophysical systematics to produce interesting predictions for upcoming experiments.\footnotemark}

~\footnotetext{This submission was written as a contribution to the 2024 cosmology session of the 58th Rencontres de Moriond.}

\section{Introduction}
Acting as a background radiation field, photons in the CMB Rayleigh Jeans tail interact with the hyperfine levels of ground state hydrogen as they propagate through the IGM.   The corresponding absorption or emission 21-cm transitions  encode crucial information of the baryonic gas and its interactions during the so called Cosmic Dawn (CD; for a review see~\cite{mesinger2019cosmic}).
Since, however, the 21-cm signal is intricately tied to unknown properties of the first astrophysical objects as the radiation they emit  interacts with neutral hydrogen, the predictions for 21-cm cosmology  notoriously suffers from significant uncertainties. Previous literature~\cite{Park:2018ljd,Fialkov:2016zyq} applied supplementary observables including high redshift UV luminosity functions (UVLF), the unresolved Cosmic X-ray Background (CXB), the optical depth to reionization, and the \Lya Ly-$\beta$ absorption lines in quasar spectra, to mitigate these uncertainties assuming minimal astrophysical modeling. However, to our knowledge, none have combined all observables necessary for constraining the entire astrophysical parameter space. In Ref.~\cite{Katz:2024ayw}, summarized in this proceedings, we combine and refine the above constraints, enabling us to bracket the evolution of the global 21-cm signal. 

The 21-cm signal can also be used   as a probe for new physics which potentially  leaves its imprint by directly producing 21-cm photons, altering the kinetic temperature of hydrogen, or affecting structure or stellar formation  (see \cite{Katz:2024ayw} and references therein). 
For this, disentangling astrophysical uncertainties from new physics effects is crucial.  With the above new constraints we demonstrate the feasibility of doing so by considering the two-fluid millicharged dark matter scenario, deriving novel prospective constraints.

\section{The Cosmic Dawn Global Signal and Astrophysical Models}\label{Sec:Modeling}




Once formed, the first astrophysical objects emit radiation that interacts with neutral hydrogen, affecting its global spin and ionization state. 
Three types are of importance:


\textbf{Lyman band emission (non-ionizing UV):} 
\Lya photons induce spin flips in neutral hydrogen through repeated resonant Raman scatterings, intricately relating the spin state of hydrogen and its kinetic temperature, $T_\text{k}$, known as the Wouthuysen Field (WF) effect~\cite{Hirata:2005mz}. These photons are produced by the redshifting and interactions of Lyman band photons, primarily sourced by the more massive (short lived) stars~\cite{Hirata:2005mz}, thus emitted with a global UV emissivity 
that traces the global star formation rate density (SFRD), $\dot{\rho}_\star$, as~\cite{Barkana:2004vb}: $\epsilon_{UV}(E,t) = {\dot{\rho}_\star(t)}\langle dN/dE\rangle/{\mu_{b}}$,
where $\mu_b$ is the average baryon mass in stars, and $\left<dN/dE\right>$ is the average number of photons emitted by a baryon in stars per unit energy.  Motivated by high redshift observations of UVLFs (see Sec.~\ref{sec:Constraints}) we model the SFRD as a power-law in halo mass with an exponential cutoff below $M_{\rm cut}$~\cite{Park:2018ljd}, accounting for the inefficient gas cooling expected in small halos. We set $\left<dN/dE\right>$ as in~\cite{Barkana:2004vb}, modeling the spectrum as
a broken power law in energy, and fitting to population synthesis models, taking the normalization of~\cite{Park:2018ljd}~\footnote{Depending on the spectra in a small energy band, the emissivity is mostly sensitive to the normalization of the spectra, which is degenerate with the normalization parameter of  $\dot{\rho}_\star$, rather than its shape.}. 



\textbf{X-ray emission:} X-ray photons in the soft $E\in [E_{\rm min}, 2\text{ keV}]$ band interact with the IGM through photoionizations, heating hydrogen. Harder photons do not interact over Hubble scale, while softer ones are absorbed in the ISM~\cite{Das:2017fys}. In the absence of high-redshift data on the ISM properties, we rely on simulations~\cite{Das:2017fys}, indicating $E_{\rm min} \in \left[0.19,0.85 \right]$~\footnote{This range corresponds to an ISM optical depth of unity for the $2\sigma$ hydrogen column density values in~\cite{Das:2017fys}.}.
Motivated by observations of nearby starburst galaxies, we consider high mass X-ray binaries as the primary soft band X-ray emitters during CD (see~\cite{mesinger2019cosmic} and references therein). Being short lived, their emissivity traces the SFRD with a power-law spectrum, $
\langle d N_X/dE\rangle/\mu_b= {5.55\times10^{-7}f_{\rm X}E^{-2}}/{\log (2\text{ keV}/E_{\rm min}} ),
$
consistent with the ISM attenuated spectrum of population synthesis models~\cite{Das:2017fys,Fragos:2013bfa}. 
Above, $f_{\rm X}$ is a free parameter that scales the X-ray luminosity to SFR ratio of HMXBs, and is expected to exceed its local value of $f_{\rm X}\sim0.1$ due to the lower metallicity at CD~\cite{Fragos:2013bfa}.

\textbf{Ionizing UV emission:} 
Typically being the dominant ionizing source, the emission of UV photons suppresses the 21-cm signal. 
We treat UV ionization following~\cite{madau1999radiative}, but with an ionization rate per baryon given by $\Gamma_{\rm{ion}} =  \frac{1}{\rho_b^0}\int \frac{d\dot{\rho}_\star}{dM_{\rm h}} N_{\rm{ion}}f_{\rm{esc}}(M_h) dM_{\rm h}$,
where $\rho_b$ is the energy density of baryons, $N_{\rm ion}$ is the number of ionizing photons emitted per baryon in stars, 
and $f_{\rm esc}$ is the fraction of ionizing photons that escape to the neutral IGM. Following~\cite{Park:2018ljd} we model the escape fraction as a power-law in halo mass, $M_{\rm h}$, having two parameters: a normalization and a power-law index. Although being simplistic, we successfully verified that the procedure above accurately matches the results of the semi numerical simulation by 21cmFAST~\cite{Park:2018ljd,Mesinger:2010ne} for $x_{HI}=\frac{n_{\rm HI}}{n_{\rm H}}>0.8$, corresponding to redshift $z \lesssim 10$.

\begin{figure}
    \centering
    \includegraphics[width=0.45\textwidth]{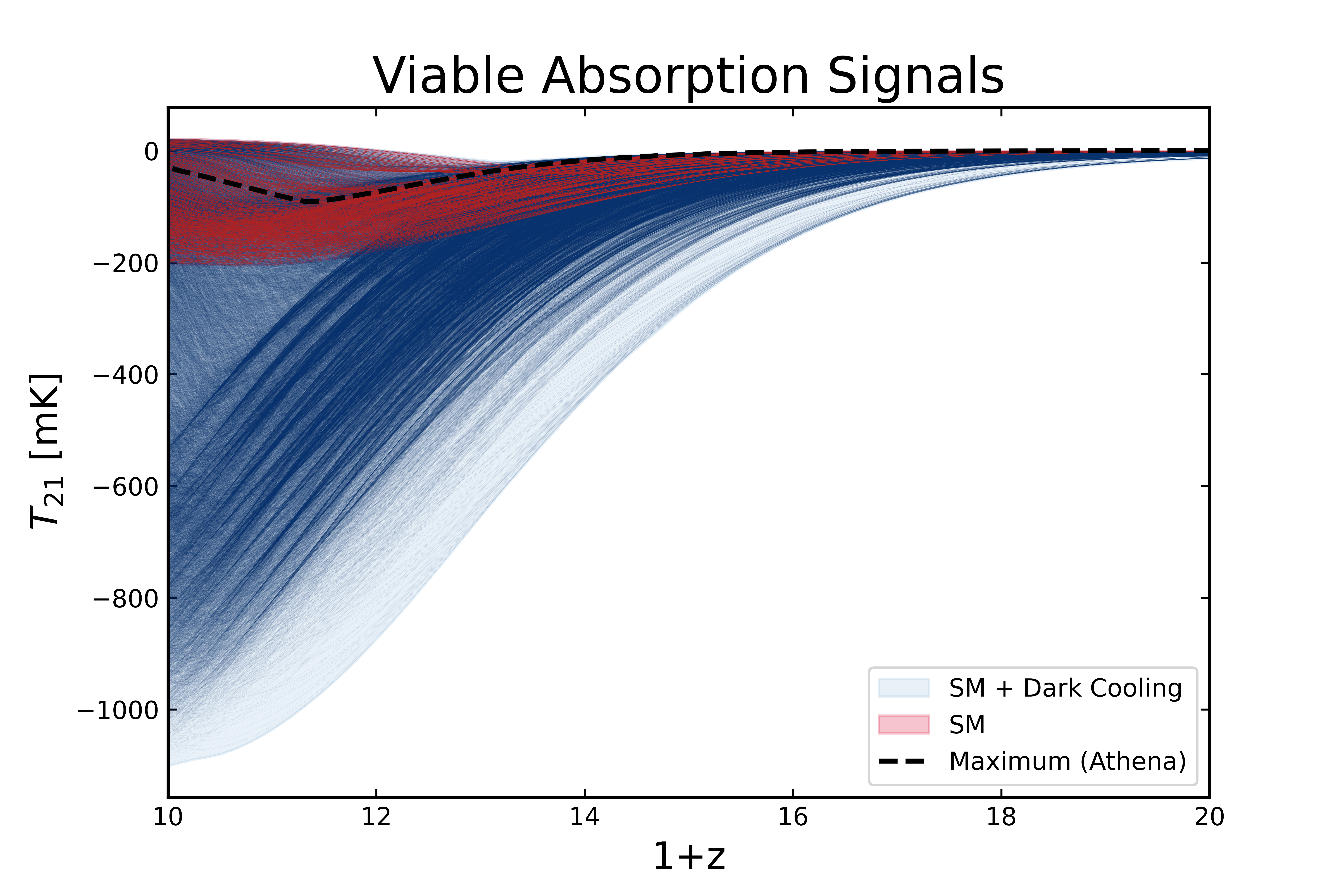}
    \includegraphics[width=0.45\textwidth]{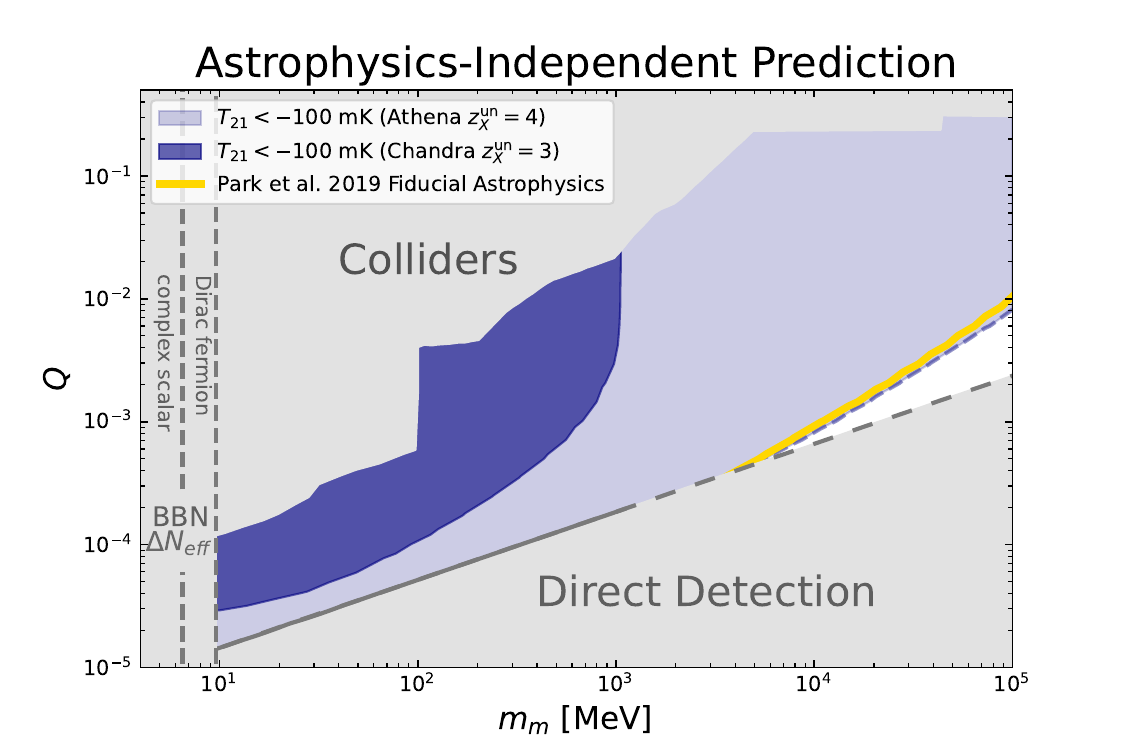}
    \caption[]{\textbf{Left:} The envelope of viable SM (2cDM dark sector) global 21-cm signals are shown in red (blue). The dashed black line is the predicted maxima of the envelopes assuming data from the future Athena X-ray survey~\cite{Marchesi:2020smf}. \textbf{Right:} For a global 21-cm measurement excluding signals below $-100\units{mK}$, our predicted constraints for the 2cDM dark sector are depicted as the dark purple region in the mDM (mass, charge) parameter space. Light shades demonstrate the expected improvement with future X-ray data from Athena. Grey shaded regions are excluded by a collection of accelerator experiments, direct detection searches, and BBN measurements of $\Delta N_{\rm eff}$,  (as further specified in~\cite{Katz:2024ayw,Liu:2019knx}).}
    \label{fig:Fig}
\end{figure}


\section{Constraining Astrophysics}\label{sec:Constraints}

The astrophysical dependence of the 21-cm signal is parameterized by three effective functions: the global UV emissivity,
the global X-ray emissivity, 
and the neutral fraction of hydrogen, $x_{\rm HI}$, which depends strongly on the ionization by UV photons.
This connection provides an opportunity to investigate the presently obscure properties of the earliest stars and galaxies. In this regard, it is interesting to bracket our predictions for the 21-cm signal using supplementary observables, enabling us to discern deviations sourced by non-standard astrophysical and cosmological phenomena, either of SM or new physics origin (see Sec.~\ref{sec:NewPhysics}). 
The main result of this section is shown as red contours in Fig.~\ref{fig:Fig} (left), covering the envelope of viable 21-cm signals, bracketed as described throughout this section. 

Applying the preliminary $2\sigma$ constraints from \cite{Park:2018ljd}, we confine the parameters associated with stellar formation, also bracketing the Lyman band emissivity for a given normalization. These constraints are derived using the $1,500$\r{A} UVLFs as a tracer for the star formation rate at redshift $6<z<10$. Relying on data taken by the Hubble Space Telescope, we expect these constraints to tighten with upcoming James Webb Space Telescope data. 





For a given SFRD, the parameters associated with X-ray heating are constrained by saturating the unresolved CXB in the $[0.5,2]\units{keV}$ band, as measured today by Chandra~\cite{Lehmer:2012ak}, with emission from the same sources that heat the IGM during CD~\cite{Fialkov:2016zyq}. Assuming all sources above redshift $z_{\rm X}^{\rm un}$ are resolved, the Chandra limit corresponds to



\begin{equation}\label{eq:Chandra}
    I_{[0.5{\rm keV}, 2{\rm keV}]}(z_{\rm X}^{\rm un}) = \frac{1}{4\pi}\int_{0.5{\rm keV}}^{2 {\rm keV}} dE E \int_{z_{\rm X}^{\rm un}}^\infty dz \frac{ \epsilon_{\rm X}(E(1+z),z)}{H(z)} e^{-\tau_{\rm X}(E,z_{\rm X}^{\rm un},z)} < 0.5 \frac{\rm keV}{{\rm cm}^2\units{sec}\units{sr}}\,,
\end{equation}
where in Fig.~\ref{fig:Fig} (left) we assume a conservative choice of $z_{\rm X}^{\rm un}=3$ compared to the highest $z\sim1$ Chandra reports of soft X-ray galaxies. Above, $\tau_{\rm X}$ is the X-ray optical depth, and $H$ is the Hubble parameter. 
$I_{[0.5{\rm keV}, 2{\rm keV}]}(z_{\rm X}^{\rm un})$ is  mostly supported by the emission around the minimal energy, corresponding to $E=0.5(1+z_{\rm X}^{\rm un})\units{keV}$, which for $z_{\rm X}^{\rm un}=3$ exceeds the upper energy boundary of our emissivity model. While previous work assumed a single power-law in all energies~\cite{Fialkov:2016zyq}, we match a second power law at $E>2\units{keV}$ with index $-3.2$, corresponding to the softest (and thus most conservative) choice allowed by population synthesis models in the energy range $[2,15]\units{keV}$~\cite{Fragos:2013bfa}. Beyond $15\units{keV}$ the exact modeling is insignificant due to the supression in the SFRD at high redshifts and in the emissivity at higher energies.

Finally, we apply constraints on the evolution of the neutral hydrogen fraction using the 68\% C.L.\ of the CMB optical depth to reionization as measured by Planck~\cite{Planck:2018vyg}.
In addition we also apply constraint from \Lya and Ly-$\beta$ absorptions in quasar spectra, which set a $1\sigma$ upper limit of $x_{\rm HI}<0.06+0.05$ at $z=5.9$~\cite{McGreer:2014qwa}. Together with the additional constraints on the SFRD and X-ray parameters, these bracket the final set of parameters associated with ionizing UV emission and ISM attenuation.



\section{Application: New Physics at Cosmic Dawn}\label{sec:NewPhysics}
To demonstrate the potential of 21-cm cosmology in the search for new physics, we assume the scenario of the two coupled DM (2cDM) dark sector~\cite{Liu:2019knx}, in which a millicharged DM (mDM) component with fraction $f_m$ resides. 
By elastically scattering with mDM, the baryonic gas in the IGM cools, thus decreasing the spin temperature of hydrogen through the WF coupling, enhancing the absorption signal at CD with respect to SM predictions as demonstrated by the blue contours in Fig.~\ref{fig:Fig}(left). The mDM interaction with the rest of the DM enhances cooling, effectively increasing the heat capacity of the small mDM fraction.

In Fig.~\ref{fig:Fig} (right) we show our predictions for constraints on the 2cDM dark sector, assuming a global signal measurement which excludes absorption signals shallower than $-100\units{mK}$ at $z\geq 10$. We project these constraints onto the 2D parameter space spanned by the millicharged mass and charge, while choosing the remaining three parameters of the model-- $f_{\rm m}$, the CDM mass, and the mDM-CDM cross section --to maximize cooling while maintaining current CMB~\cite{Boddy:2018wzy} and BBN constraints~\cite{Boehm:2013jpa}. In the absence of high redshift astrophysical data, significant systematics hinder the constraints.
For instance, an increased X-ray heating, suppressed Lyman band emission or a higher ionizing UV emission will lead to a shallower signal by increasing $T_{\rm k}$, weakening the WF coupling, or decreasing the abundance of neutral hydrogen. To remove the astrophysical systematics from our predicitions, for each choice of mDM mass and charge we scan over the entire viable astrophysics parameter space, which is determined according to the procedure described in the previous section. We then consider a choice of mDM parameters as constrained only if $T_{21}<-100\units{mK}$ for all viable astrophysical models.

Our predictions demonstrate the crucial role that the 21-cm global signal can play in probing new physics, specifically here probing the 2cDM dark sector. Our predictions extend beyond the reach of current collider and direct detection experiments, having the possibility to probe a large region of the viable parameter space. We expect these predictions to improve significantly with the future X-ray survey by Athena~\cite{Marchesi:2020smf} as we show in Fig.~\ref{fig:Fig}.


\section*{Acknowledgments}
OZK thanks the Alexander Zaks scholarship for its support.

\section*{References}

\bibliography{Moriond.bib}






\end{document}